# A consistent model for cardiac deformation estimation under abnormal ventricular muscle conditions

Ahmad. R. Baghaie[1], H. Abrishami Moghaddam[2]

[1] K.N.Toosi University of Technology/ Biomedical Engineering, M.Sc. Candidate, Tehran, Iran
[2] K.N.Toosi University of Technology/ Biomedical Engineering, Associate Professor, Tehran, Iran

*Abstract*— **Deformation modeling of cardiac muscle is an important issue in the field of cardiac analysis. For this reason, many approaches have been developed to best estimate the cardiac muscle deformation, and to obtain a practical model to use in diagnostic procedures. In this paper, using a point-wise approach in deformation estimation, we try to estimate the deformation under some abnormal conditions of cardiac muscle. First, the endocardial and epicardial contour points are ordered with respect to the center of gravity of endocardial contour and displacement field is extracted. Then to solve the governing equation in deformation, we apply boundary conditions in accordance with computed displacement field. Using obtained displacement field through the cardiac muscle, strain map is extracted to show the mechanical behavior of cardiac muscle.**

**The proposed algorithm was implemented in MATLAB and the results of the algorithm on a non-homogenous ring model showed a good correlation with the results obtained by ANSYS.**

## I. INTRODUCTION

The development of consistent models for both accurately estimating cardiac deformations and taking into account organ's regional abnormalities make this issue an interesting field. On the other hand, due to the diagnostic importance of such analysis and the inherent restrictions of medical instruments in data acquisition, new approaches are proposed to facilitate the interpretation of the acquired data. Low contrast and noisy cardiac images provided by medical imaging systems are examples of the restrictions in biomedical devices. To overcome the above limitations, medical data analysis methods use a priori knowledge about organ's behavior. These model-based approaches can achieve more accurate and robust results. In the model based analysis, using natural properties of the organ, we try to improve the performance of algorithms.

Inerney and Terzopoulos proposed a deformable dynamic balloon [1]. In their model a balloon deformes until being matched with endocardial contour in CT scan images. Pham et al. used an active region model for segmentation and tracking of cardiac deformation in heart MRI images [2]. Active contour model is used in [3] for left ventricular surface extraction. In [4], Mogaddam et al. utilized a deformable snake model for dynamic visualization of left ventricle through one cardiac cycle. However none of the above methods has the point-wise tracking ability. Mesh based approaches were introduced to overcome this problem. Nakaya et al. used triangle mesh for motion estimation [5]. Wang et al. utilized rectangle mesh to cardiac image feature extraction [6]. In [7] Malassiotis et al. proposed a method for non-rigid body deformation using triangle Delaunay mesh. In their approach, deformation approximation is based on gray level similarity between points in two consequent frames. Mosayebi et al. in [8] used 3D active mesh and tracked deformation of cardiac muscle. They first derive an initial surface of the heart and associate a 3D mesh to it, then by minimizing the strain cost function, track the cardiac deformation [8]. Jamali et al. improved this approach by optimizing the segmentation process using level-set method [9,10]. Kermani et al. in [14], estimate the local and global left ventricular function by fitting three-dimensional active mesh model (3D-AMM) to the initial sparse displacement which is measured from an establishing point correspondence procedure.

However in all of the previous methods, the inherent assumption is that the cardiac muscle is in normal condition. In other words, all of the above models consider the cardiac muscle with no abnormality which is not true in most cases. In this paper, using a point-wise 2D manner, we try to model the deformation of cardiac muscle under some abnormal conditions. In this case, the ventricular muscle may contains some inactive regions which cause different behavior of muscle during the cardiac cycle. In this regard, the whole cardiac motion can be affected by loading of these regions.

The paper is organized in this way: We introduce our proposed model in chapter II. In chapter III, the proposed algorithm will be implemented using MATLAB and then, we compare the resultat with the same problem analysed by ANSYS. Generalized algorithm for real left cardiac muscle extracted from MRI images and its results for strain are given at the end of chapter III. The discussion and conclusion come in chapter IV.

## II. PROPOSED MODEL

### A. Displacement vector definition

For deformation modeling of a deformable body, it is necessary to know the displacement field through the body. So, assume that $X(0)$ describes the coordinates of one point in $t=0$ and $X(t_0)$ describes the coordinates of the same point in $t=t_0$. So the displacement vector between these two moments can be defined [11]:

$$U = X(t_0) - X(0) \qquad (1)$$

In this manner, knowing the position of all body points in two initial and deformed stages, the displacement field can completely be computed. However, finding a one-to-one correspondence for each point between these two stages is a challenging work. Especially when the number of points increases, this procedure becomes more complicated. On the other hand, due to nonlinear nature of deformation in deformable organs of human body, tracking of all the points during the deformation is almost impractical.

There are some works which try to simplify this procedure. In [15], Shi et al. utilized a curvature parameter to find the correspondence between points in two different stages. In this paper, we propose a more efficient and simpler method for finding corresponding points between two frames.

Assume that $I(0)$ and $O(0)$ represent inner and outer contours of a deformable body in $t=0$ and $I(t_0)$ and $O(t_0)$ represent inner and outer contours in $t=t_0$, respectively. Also assume $[x_c, y_c]$ represents the origin of the coordinates system; hence all contour points are represented with respect to the origin. More precisely, the coordinates of the origin don't change between two frames and it can be supposed as basis point in two times. Therefore, as a first step, the reference point (origin) should be defined.

### B. Reference point definition

Definition of the reference point is a critical step in this algorithm; since all the contour points will be ordered in accordance with their respective positions around the reference point and the displacement field is defined based on the ordered contour points. Therefore, the reference point should be defined more consistently with the deformation. However, there are some facts in the deformation of the left ventricle which can help us to define the reference point more efficiently.

- Generally, the deformation of the left ventricle is approximately concentric. In other words, we can assume that all the contour points act like a system of mass and spring around a center between two frames.

- Most of the researches reported on modeling and tracking of cardiac muscle suppose that the deformation of cardiac body between two frames takes place satisfying small strain deformation condition. This assumption is aproximately true in such case and obtained results prove it.

Based on the above assumptions, the reference point is defined as follows:

$$[x_c, y_c] = \frac{1}{N}\sum_{i=1}^{n}[x_i, y_i] \qquad (2)$$

where $N$ is the number of inner contour points and $[x_i, y_i]$ represent the coordinates of $i$'th contour point. It is worthy to note that this definition is appropriate only for convex or semi-convex contours such as cardiac wall boundaries. Finally, all the contour points are numbered anti-clockwise starting from the point positioned at the right hand side and horizontally with respect to the reference point.

### C. Governing equation solution using Finite Element Method

Suppose that $\Omega$ is the domain of deformable body and $\partial\Omega$ is the boundary of the domain. Generally, governing equation for a deformable body is an elliptic equation as follows [12]:

$$-\nabla(c(u).\nabla u) + a(u)u = f(u) \quad on \quad \Omega \qquad (3)$$

where $c$, $a$, and $f$ are functions of the unknown solution $u$ in domain $\Omega$. Such equation can be solved using the following Dirichlet and Generalized Neumann boundary conditions, respectively:

$$
\begin{aligned}
hu &= r \quad on \quad \partial\Omega \\
\vec{n}.(c\nabla u) + qu &= g \quad on \quad \partial\Omega
\end{aligned}
\qquad (4)
$$

where $\vec{n}$ is the outward unit normal vector and $g$, $q$, $h$, and $r$ are complex valued functions defined on $\partial\Omega$. For the two-dimensional domain, the Dirichlet boundary conditions are:

$$
\begin{aligned}
h_{11}u_1 + h_{12}u_2 &= r_1 \\
h_{21}u_1 + h_{22}u_2 &= r_2
\end{aligned}
\qquad (5)
$$

In this paper, the Dirichlet boundary condition is used based on computed displacement values. Then, the governing function is computed using FEM analysis [12].

## III. SIMULATION AND VALIDATION

### A. Modeling a non-homogeneous ring in ANSYS

Modeling the left ventricle wall deformation during the cardiac cycle is a challenging work. In addition, in this paper we want to model a non-homogeneous cardiac muscle. So the modeling procedure is more restricted.

In the case of plane stress, the model can be defined by a constant thickness cylinder. In this manner, the deformation of body along $Z$ axis is eliminated. Also the difference in mechanical parameters in cardiac muscle, caused by myocardial infarction is modeled using 2 merged areas in ANSYS. The mechanical parameters of cylinder, E1 for Young modulus and v1 for Poisson's ratio for normal and E2 and v2 for abnormal region are set as 31000, 0.45, 310000 and 0.45, respectively. In loading stage, the cylinder is affected by a pressure inside it. Fig. 1 shows the magnitude of displacement of ring caused by applied pressure.

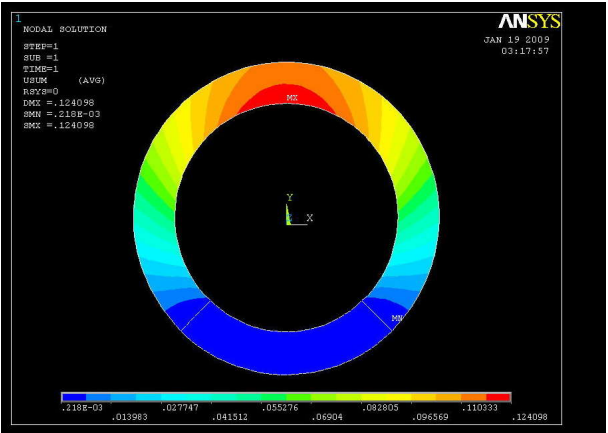

Fig. 1. Magnitude of displacement of ring

### B. Modeling the non-homogeneous ring in MATLAB

Modeling of non-homogeneous ring then take places in MATLAB. For this, at first inner and outer contours of ring, in both states (initial and deformed) will be extracted using special procedures for image labeling. Based on the above procedure for ordering the contour points with respect to the reference point, the points are numbered and then, displacement field will be extracted and used as boundary conditions to solve the governing equation by FEM.

For more accurate analysis of the results, it is more appropriate to split the ring in parts. In this regard, the ring is divided to 16 sections. Then, computed displacement fields along X and Y axis can be compared between ANSYS and MATLAB model. Table. 1 shows the correlation of the computed displacement through the cylinder between ANSYS and MATLAB.

### C. Modeling the cardiac muscle deformation

For modeling and analysis of mechanical parameters of cardiac muscle, cardiac MRI image database in [13] is used. This database contains MRI image sets for 33 subjects.

Each data set consists 8 to 15 slices of the heart muscle, there are 20 short axis image for each slice, representing the deformation of slice during the cardiac cycle starting from the beginning of systole to end of diastole. There is 6 to 13 mm distance between two slices. Endocardial and epicardial contours of the left ventricle in images are manually segmented by the author of [13] and each contour is represented with 32 points. For this reason, at first we interpolate between the contour points to achieve higher resolution. Then, to reduce the computational time, the contour points were uniformly down sampled by factor 9.

Then the displacement fields through the deformed wall can be computed using the method described in section II, part $A$. Now, suppose that $u$ and $v$ are displacement functions along $X$ and $Y$ axis, respectively. Then strain components can be computed using [12]:

$$\begin{bmatrix} \varepsilon_x \\ \varepsilon_y \\ \gamma_{xy} \end{bmatrix} = \begin{bmatrix} \dfrac{\partial}{\partial x} & 0 \\ 0 & \dfrac{\partial}{\partial y} \\ \dfrac{\partial}{\partial y} & \dfrac{\partial}{\partial x} \end{bmatrix} \begin{Bmatrix} u \\ v \end{Bmatrix} = [\partial]\{u\} \qquad (6)$$

where $\varepsilon_x$, $\varepsilon_y$ and $\gamma_{xy}$ represent strain along $X$, $Y$ axis and shear strain, respectively. Fig. 2 illustrates strain components of cardiac muscle between two frames of cardiac cycle.

## IV. CONCLUSION AND DISCUSSION

We have proposed a reliable model for tracking and extracting of mechanical properties of abnormal cardiac muscles. Previous models always supposed that the cardiac muscle is an isotropic homogeneous material which in most cases is not true. At first, an algorithm for fitting 2D cardiac contours during the cardiac cycle is proposed. Then using the definition of displacement vector, we extract displacement field on the left ventricle wall. Based on a FEM based algorithm, we identified the displacement field through the wall. Then, strain components were computed using continuum mechanics basics.

For validation of the proposed method, first we modeled a non-homogeneous ring in ANSYS and extracted displacement field through the deformable body. Then the displacement field within the wall was computed using MATLAB. Comparing ANSYS model and MATLAB proposed method shows a high correlation between the responses in two environments. After validation of the proposed model, we tested our algorithm on real cardiac MRI images. The proposed method is more accurate model and can be used in both normal and abnormal cardiac muscle conditions.

Table 1 Correlation of computed radial displacement of non-homogeneous ring for 16 sections (each section 22.5 degrees) between ANSYS and MATLAB.

| Sections | 1 | 2 | 3 | 4 | 5 | 6 | 7 | 8 | 9 | 10 | 11 | 12 | 13 | 14 | 15 | 16 |
|---|---|---|---|---|---|---|---|---|---|---|---|---|---|---|---|---|
| Correlation | .9803 | .9990 | .9995 | .9999 | 1.000 | .9996 | .9994 | .9981 | .9960 | .9932 | .9981 | .9766 | .9913 | .9846 | .9804 | .9731 |

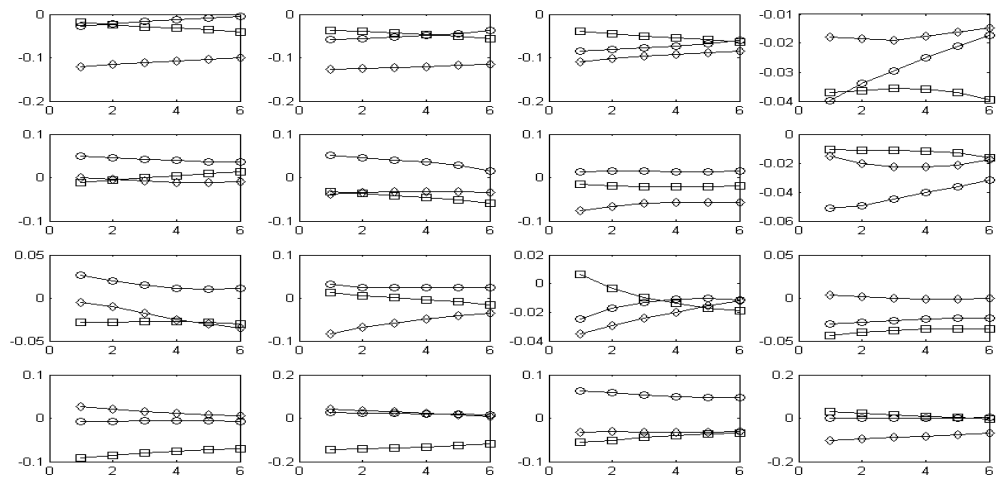

Fig. 2 Strain components in 16 sections of cardiac muscle (left to right, from the top). Diamond shaped: X direction strain, Square shaped: Y direction strain, Circle shaped: Shear strain.

Author:    H. Abrishami Moghaddam
Institute:  K.N. Toosi University of Technology
Street:     Shariati St., Seyed Khandan
City:       Tehran
Country:   Iran
Email:     moghadam@eetd.kntu.ac.ir